\documentclass[aps,prl, twocolumn,groupedaddress,showpacs]{revtex4-1}

\usepackage{graphicx}
\usepackage[geometry]{ifsym}
\usepackage{multirow}
\usepackage{dcolumn}
\usepackage{MnSymbol}

\begin{document}
%\preprint{}


\title{Universal dissipation scaling for non-equilibrium turbulence}



%\author{J. C. Vassilicos and P. C. Valente}
\author{P. C. Valente}
\email[]{p.valente09@imperial.ac.uk}
\thanks{}

\author{J. C. Vassilicos}
\email[]{j.c.vassilicos@imperial.ac.uk}
\affiliation{Turbulence, Mixing and Flow Control Group, Department of Aeronautics, Imperial College London}
\homepage[]{www.imperial.ac.uk/tmfc}

\date{\today}

\begin{abstract}
It is experimentally shown that the non-classical high Reynolds number energy dissipation behaviour, $C_{\varepsilon}\equiv \varepsilon L/u^3 = f(Re_M)/Re_{L}$, observed during the decay of fractal square grid-generated turbulence is also manifested in decaying turbulence originating from various regular grids. For sufficiently high  values of the global Reynolds numbers $Re_M$,  $f(Re_M)\sim Re_M$.

\end{abstract}


\maketitle

%%%%%%%%%%%%%%%%%%%%%%%%%%%%%%%%%%%%
% paragraph 1
In recent papers describing the wind tunnel turbulence generated by
fractal square grids \cite{VV11,MV10} it was shown that the turbulent kinetic energy dissipation rate, $\varepsilon$, at moderately high Reynolds
numbers does not follow the expected scaling $\varepsilon L/u^{3} \equiv
C_{\varepsilon} \approx const$ (where $L$ is the longitudinal integral length-scale and $u$ the streamwise r.m.s. velocity). Instead \cite{VV11,MV10} found that $C_{\varepsilon} = f(Re_M)/Re_{L}$ during the turbulence decay where $f(Re_{M})$ is an increasing function of $Re_{M} = U_{\infty} M/\nu$, a global Reynolds number based on a length-scale $M$ characteristic of the grid, and where $Re_{L} = u L
/\nu$ is a local, downstream position dependent, Reynolds number
($\nu$ is the kinematic viscosity and $U_{\infty}$ is the inflow
velocity). This behaviour is accompanied by a well-defined power-law
energy spectrum (with exponent close to Kolmogorov's $-5/3$) over a
broad range of length-scales and is therefore caused by a physically different underlying phenomenon than the well-known low Reynolds number law $C_{\varepsilon}\sim Re_{L}^{-1}$.

% paragraph 2
Evidence of such a non-classical behaviour is significant due to the central role the empirical law $C_{\varepsilon}\approx const$ has on most, if not all, models and theories of both homogeneous and inhomogeneous turbulence \cite{Townsend:book,TennekesLumley:book,MY:book,Frisch:book}. Clearly, one should expect the existing models to inadequately describe turbulent flows (or regions thereof) not obeying the $C_{\varepsilon}\approx const$ scaling and consequently fail in their predictions of transport phenomena (energy transfer, dissipation, particle dispersion, scalar diffusion, etc...). Most importantly, it challenges our understanding of turbulence phenomena in general, nevertheless providing a starting point for its study as well. 

% paragraph 3
In this Letter we report results which show that this non-classical behaviour is in fact more general than previously thought and is not exceptional to the very special class of inflow conditions defined by fractal square grids. Hence this non-classical behaviour is of general scientific and engineering significance and therefore of much greater importance. 

\begin{figure}
\centering
\includegraphics[width=85mm]{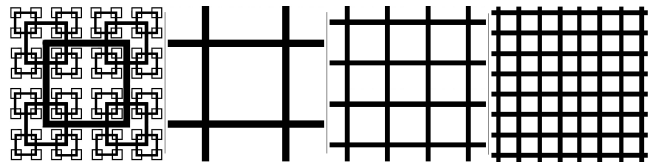}
\caption{\label{figGrids} Turbulence generating grids. From left to right: FSG\cite{VV11}, RG230, RG115 and RG60.}
\end{figure}

% paragraph 4
In the present experiments we compare turbulence generated by three different regular square-mesh grids (RG230, RG115 and RG60) with the turbulence generated by the fractal square grid (FSG) of \cite{VV11} (see Fig. \ref{figGrids} and table \ref{grids}). Our aim is to investigate the origin for the non-classical dissipation behaviour of the FSGs. The dimensions of RG230 are purposefully similar to those of the largest square on the FSG. This allows a \emph{ceteris paribus} comparison between RG230 and FSG in two respects: (i) comparable inflow Reynolds numbers $Re_M$ for similar inflow velocities if $M$ is taken to be the side-length of the largest square on the grid (see Fig. \ref{figGrids}) and (ii) comparable distance from the grid where the wakes of the RG230 bars meet and where the wakes of the FSG largest bars meet. Starting from any one of our grids, the turbulent kinetic energy increases as one moves downstream along the tunnel's centreline and reaches a peak at a streamwise distance $x_{peak}$ from the grid beyond which the turbulence decays \cite{VV11,MV10,J&W1992}. This distance $x_{peak}$ is closely related to the distance from the grid where the wakes (largest wakes in the case of FSG) meet. Indeed, \cite{MV10} introduced the wake interaction length-scale $x_{*} \equiv M^{2}/t_{0}$ where $t_0$ is the lateral thickness of the largest bars (see Fig. \ref{figGrids}) and showed that $x_{peak}$ scales with $x_*$ in the case of FSGs. Subsequently, \cite{VV11} showed that $x_{peak}/x_{*}$ took comparable values for RGs and SFGs, a point which the experiments reported in this Letter allow us to confirm (see table \ref{grids}). The length-scales $x_{peak}$ and $x_{*}$ turn out to be paramount for a meaningful comparison between grids.

% paragraph 5
There are of course important differences between the four grids used here, for example different values of blockage ratio $\sigma$ (ratio between the blocking area of the grid and the area of the tunnel's test section) and different values of $x_{*}$ (see table \ref{grids}). These differences cause differences in various mean flow and turbulence profiles across the tunnel section. However, they have no baring on our main finding that the outstanding behaviour previously found in FSG-generated turbulence is also present in turbulence generated by regular grids for a region whose extent is determined by $x_*$. Beyond this region, in the one case (RG60) where we can reach sufficiently far beyond it as a result of the wind tunnel's test section being much longer than $x_*$, we find the classical behaviour $C_{\varepsilon} \approx const$ provided the Reynolds number is sufficiently high.

% paragraph 6
The experimental apparatus described in \cite{VV11} was repeated for the present experiments with the length of the 0.46m x 0.46m-wide test section shortened from $\approx4.5\mathrm{m}$ to $\approx 3.5\mathrm{m}$ to match the extent of the longitudinal traverse mechanism. We also installed a grid at the entrance of the diffuser to maintain a slight overpressure across the test section.
 All data are recorded with one- and two-component hot-wire anemometers operated at constant temperature. The main data are recorded with two in-house etched Pl-(10\%)Rh single-wire ($SW$) sensors, $SW1$ and $SW2$, having sensing lengths of $l_w=0.5\mathrm{mm}$ and $l_w=0.2\mathrm{mm}$ and wire diameters of $d_w=2.5\mu\mathrm{m}$ and $d_w=1\mu\mathrm{m}$, respectively. A Dantec 55P51 cross-wire (XW) with $l_w=1.0\mathrm{mm}$ and $d_w=5\mu\mathrm{m}$ is also used to record basic isotropy statistics. 
The spatial resolution of the measurements, quantified by $l_w/\eta$ ($\eta\equiv \left(\nu^3/\varepsilon \right)^{1/4}$ is the Kolmogorov microscale; the isotropic estimate of dissipation $\varepsilon=15\nu\overline{(du/dx)^2}$ is used), is given  in table \ref{Table:Results} for the furthermost up- and downstream locations and for the different inflow velocities.  
We repeated the electronic tests to confirm that the maximum unattenuated frequency response of the $SW$s was at least $k\eta = 1$ ($k$ is the wavenumber). 
The data acquisition and processing methodologies are also similar to those described in \cite{VV11}. An exception is that we use, for simplicity, the classical  Taylor's frozen field hypothesis to convert temporal into spatially varying signals, although we checked that this does not meaningfully affect the results. 

% paragraph 7
This Letter's new data are recorded along the centreline in the lee of each of  our four grids (Fig. \ref{figGrids} and tables \ref{grids} and \ref{Table:Results}). 
Data recorded between a grid and its corresponding $x_{peak}$ are excluded (see caption of table \ref{Table:Results}) as we confine our study to decaying turbulence.
In these decay regions, $u/v$ (where $v$ is the r.m.s. lateral velocity) is typically between $1.2$ and $1.1$ and the ratio of the mean square of the lateral
turbulence velocity derivative with respect to the streamwise coordinate $x$ to the mean square of the streamwise turbulence velocity derivative with respect to $x$ takes values between $1.4$ and $1.6$. 
Both ratios vary by less than $5\%$ along the streamwise extent of our records. Note that $x_{peak}$ is about as long as half the wind tunnel's extent in the cases of RG230 and FSG (see table \ref{grids}). The RG60 was investigated in \cite{VV11} where it was shown that for sufficiently high inlet velocities the dissipation followed a convincing $C_{\varepsilon}\approx const$ during decay far downstream. We repeat those measurements using a higher resolution sensor ($SW2$) and include recordings much closer to the grid (table \ref{Table:Results}). 

\begin{table}
\caption{\label{grids} Details of turbulence-generating grids; $d$ is the longitudinal thickness of the bars.}
\begin{ruledtabular}
\begin{tabular}{lccccccc}
Grid &  & $M$ & $t_0$ & $d$ & $\sigma$ & $x_*$ & $x_{peak}/x_*$ \\
        &  & (mm) & (mm) & (mm) & (\%) & (m)&  \\
\hline
RG230 & mono-planar & 230    & 20    & 6   & 17 & 2.65 & 0.63\\
RG115 & mono-planar & 115    & 10    & 3.2   & 17 & 1.32& 0.63 \\
RG60   & bi-planar       & 60      & 10    & 10 & 32 & 0.36 & $\simeq 0.4$\footnote{Taken from measurements of a very similar grid. }\\
FSG     & mono-planar & 237.7 & 19.2 & 5   & 25 & 2.94 & 0.45\\ 
\end{tabular}
\end{ruledtabular}
\end{table}

\begin{table}
\caption{\label{Table:Results} Overview of the experimental results. $x_{min}$ \& $x_{max}$ are the first and last measurement locations corresponding to  $0.48 x_*$ \& $1.09 x_*$, $0.64 x_*$ \& $1.19 x_*$, $0.61 x_*$ \& $2.38 x_*$ and $0.72 x_*$ \& $8.75 x_*$ for FSG, RG230, RG115 and RG60, respectively. Probe $SW1$ is used for the measurements of the first two grids and $SW2$ for the last two.}
\begin{ruledtabular}
\begin{tabular}{lcccccc}
Grid & Symbol & $U_{\infty}$ & $Re_{M}$& $u/U_{\infty}(\%)$ & $Re_{\lambda}$ & $l_w/\eta$ \\ \cline{5-7}
 &&$\left(\mathrm{ms}^{-1}\right)$& $\left(\times 10^{3}\right)$& \multicolumn{3}{c}{$x_{min}$ / $x_{max}$} \\
\hline
\multirow{2}{*}{FSG} &\rlap{\SmallSquare}\SmallCross& $15.0$ & $237$ & \multirow{2}{*}{$9.7 \,/\, 5.0$}  & $385 \,/\, 249$  & $4.8\,/\, 3.0$  \\
&\rlap{\SmallCircle}\SmallCross& $17.5$ & $277$ && $418 \,/\, 275$ & $5.5 \,/\, 3.5$ \\ \\

\multirow{5}{*}{RG230} & \FilledDiamondshape & $5.0$ & $77$ & \multirow{5}{*}{$7.2\,/\, 4.8$} & $180 \,/\, 140$ & $1.8 \,/\, 1.3$  \\
&\SmallSquare& $10.0$ & $153$ && $261 \,/\, 200$ & $2.9 \,/\, 2.2$ \\
&\FilledSmallCircle& $15.0$ & $230$ && $326 \,/\, 258$ & $3.9 \,/\, 3.0$ \\
&\Diamondshape& $17.5$ & $268$ && $348 \,/\, 281$ & $4.4 \,/\, 3.3$ \\
&\Large{$\filledstar$}& $20.0$ & $307$ && $385 \,/\, 300$ & $4.9 \,/\, 3.7$  \\  \\

\multirow{1}{*}{RG115} &\scriptsize{\rlap{\Circle}\FilledDiamondshape}& $20.0$ & $153$ & \multirow{1}{*}{$6.9 \,/\, 2.7$}  & $255 \,/\, 160$  & $2.3\,/\, 1.1$  \\ \\

\multirow{3}{*}{RG60} & \SmallTriangleLeft& $10.0$ & $40$  & \multirow{3}{*}{$15 \,/\,  2.2$}  & $177 \,/\, 96$ & $2.8 \,/\, 0.6$  \\
&\FilledSmallTriangleUp& $15.0$ & $60$ && $240 \,/\, 111$ & $3.8 \,/\, 0.8$ \\
&\SmallTriangleRight & $20.0$ & $80$ && $290 \,/\, 135$ & $4.7 \,/\, 1.0$  \\
\end{tabular}
\end{ruledtabular}
\end{table}

%figure 1

%figure 2
\begin{figure}
\centering
\includegraphics[trim = 45 1 20 17, clip=true,width=90mm,height=62mm]{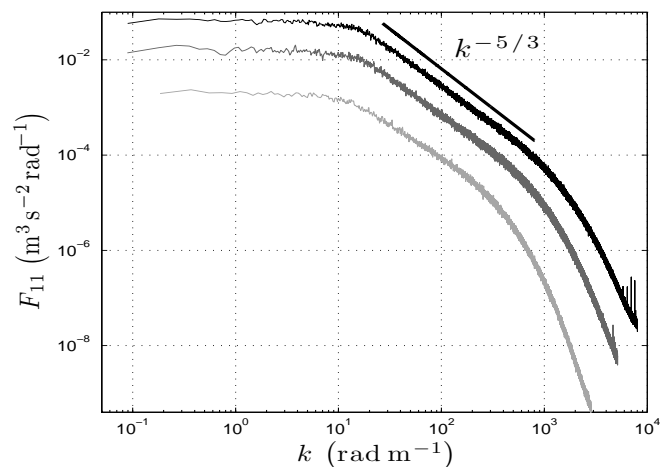}
\caption{\label{fig0} Longitudinal energy density spectra $F_{11}$ per wavenumber $k$ of turbulence generated by RG230 for (black) $U_{\infty}=20 \mathrm{m}\mathrm{s}^{-1}$, $x/x_*=0.64$, (dark grey) $U_{\infty}=10 \mathrm{m}\mathrm{s}^{-1}$, $x/x_*=0.64$ and (light grey) $U_{\infty}=5 \mathrm{m}\mathrm{s}^{-1}$, $x/x_*=1.19$.}
\end{figure}
 
\begin{figure}
\centering
\includegraphics[trim = 17 5 20 17, clip=true,width=90mm,height=62mm]{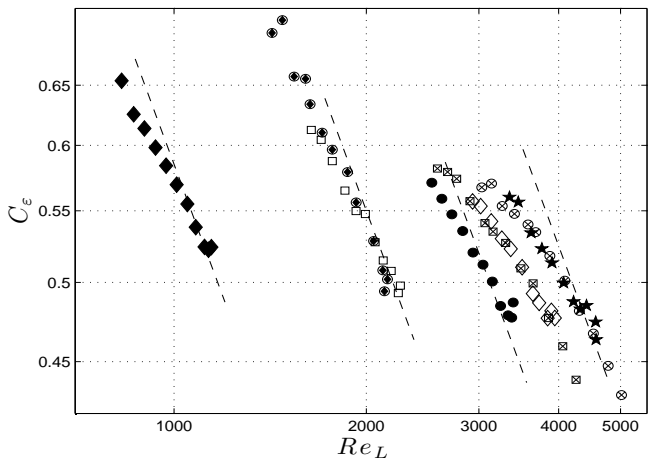} %trim=l b r t
\caption{\label{fig1} Normalised energy dissipation $C_{\varepsilon}$ versus local Reynolds number $Re_{L}$ of turbulence generated by FSG, RG230 \& RG115  for different inflow Reynolds numbers $Re_M$. The dashed lines follow $\propto \!\! Re_{L}^{-1}$ for different $Re_M$. The $Re_{\lambda}$ values of the data in this plot range between 140 and 418.}
 \end{figure}

 % paragraph 8
First, we compare the dissipation scalings of the decaying turbulence originating from RG230 and FSG. The Reynolds numbers $Re_{\lambda} \equiv u\lambda /\nu$ (where $\lambda$ is the Taylor microscale) at our measurement stations are given in table \ref{Table:Results} and are all large enough for a significant separation to exist between the large, energy containing, eddies and the smallest dissipative eddies. Indeed, the scale separation at the highest Reynolds number is $L/\eta
\approx 460$. 
The measured one-dimensional longitudinal energy spectra
$F_{11}$ exhibit clear power-laws over more than a decade with an
exponent close to Kolmogorov's $-5/3$, at least for $Re_M \ge 2.3\! \times\! 10^{5}$ and $Re_{\lambda} \ge 250$ (see Fig. \ref{fig0} where we only plot RG230
spectra for brevity and clarity; FSG spectra can be found in \cite{VV11}). 
However, both for RG230 and SFG, the cornerstone assumption of turbulence theory, $C_{\varepsilon} \approx const$, does not hold in this region where the turbulence decays (between about 1.3m from the grid and the end of the test section) at these Reynolds numbers (see Fig. \ref{fig1}). 
Instead, for any fixed $Re_M$, $C_{\varepsilon} \sim Re_{L}^{-1}$ (as one moves along $x$) is a good qualitative approximation (in Fig.  \ref{fig1} each set of symbols corresponds to one $Re_M$ and one grid, see table  \ref{Table:Results}; $Re_{L}$ decreases as $x$ increases). 
At the furthest downstream locations which correspond to the lowest $Re_L$ values for each $Re_M$ in Fig. \ref{fig1}, there is a slight departure from $C_{\varepsilon} \sim Re_{L}^{-1}$, probably due to far downstream test section confinement effects discussed in \cite{VV11}.
(In our records, $L$ reaches a maximum value smaller than $M/4$ at $x_{max}$ for all grids.)
Note that the well-known relation $\varepsilon = 15 \nu u^{2}/\lambda^{2}$ (e.g. \cite{TennekesLumley:book}) and the definition of $C_{\varepsilon}$ imply $15 (L/\lambda)^{2} = C_{\varepsilon} Re_{L}$ and $15 L/\lambda = C_{\varepsilon} Re_{\lambda}$ which means that $C_{\varepsilon} \sim Re_{L}^{-1}$ is equivalent to $C_{\varepsilon} \sim Re_{\lambda}^{-1}$ and that such $C_{\varepsilon}$ behaviour implies $L/\lambda \approx const$ during decay.  

% figure 3
 
%figure 4
\begin{figure}
\centering
\includegraphics[trim = 17 5 20 21, clip=true,width=90mm,height=63mm]{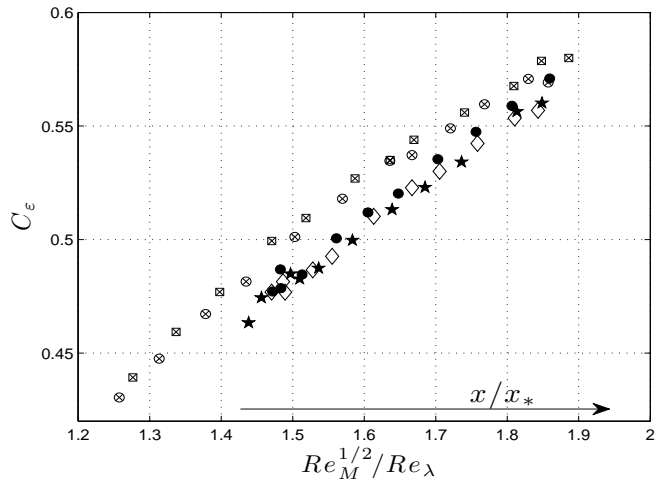}
\caption{\label{fig2} Normalised energy dissipation $C_{\varepsilon}$ versus the Reynolds number ratio $Re_M^{1/2}/Re_{\lambda}$ of turbulence generated by RG230 and FSG  for different inflow Reynolds numbers $Re_M$. }
\end{figure}
 
 % paragraph 9
When, instead of keeping $Re_M$ fixed and varying $x$, we keep $x$
fixed and vary $Re_M$, we then find a very different dependence of
$C_{\varepsilon}$ on Reynolds number; asymptotically independent of it
for both RG230 and FSG as $Re_M$ increases. 
If we keep with the usual expectation that $C_{\varepsilon}$ is independent of $\nu$ at high enough $Re_M$ (which may be close to, but not exactly, true, see \cite{M&V2008}), then these two different dependencies on Reynolds number can be reconciled by
\begin{equation}
C_{\varepsilon} \propto \frac{Re_M}{Re_L} \propto \frac{Re_M^{1/2}}{Re_{\lambda}}
\label{eq1}
\end{equation}
because $u/U_{\infty}$ and $L/M$ are independent of $Re_M$ to leading order at high enough Reynolds numbers. Note that $C_{\varepsilon}\! \propto\! Re_{M}/Re_{L}$ is equivalent to $L/\lambda \sim Re_{M}^{1/2}$ and therefore to $C_{\varepsilon}\! \propto\! Re_{M}^{1/2}/Re_{\lambda}$.
This equation is fairly well supported by our data both for FSG and RG230 at $Re_{M} \ge 2.3\!\times\!10^{5}$  (Fig. \ref{fig2}) but with a grid-dependent constant of proportionality in (\ref{eq1}). 

%figure 5
\begin{figure}
\centering
\includegraphics[trim = 15 6 20 17, clip=true,width=90mm,height=62mm]{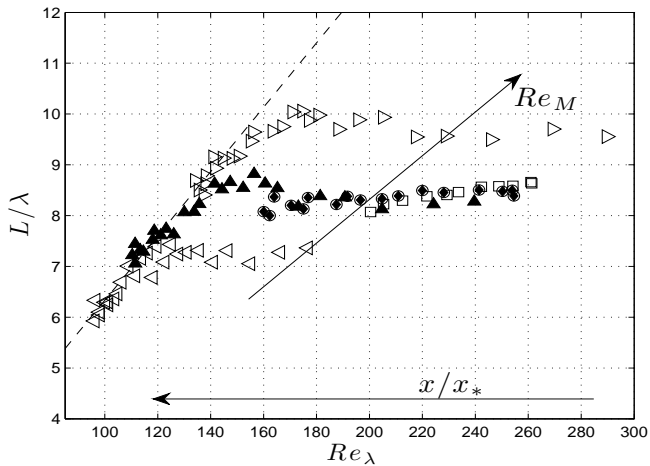}
\caption{\label{fig3}  $L/\lambda$ versus the local Reynolds number $Re_{\lambda}$ of turbulence generated by the RG60 for different $Re_M$ and by RG115 \& RG230 for the same $Re_M$. The dashed line follows $C_{\varepsilon}/15\, Re_{\lambda}$ with $C_{\varepsilon}=0.92$.}
\end{figure}

%figure 6
\begin{figure}
\centering
\includegraphics[trim = 15 6 20 17, clip=true,width=90mm,height=62mm]{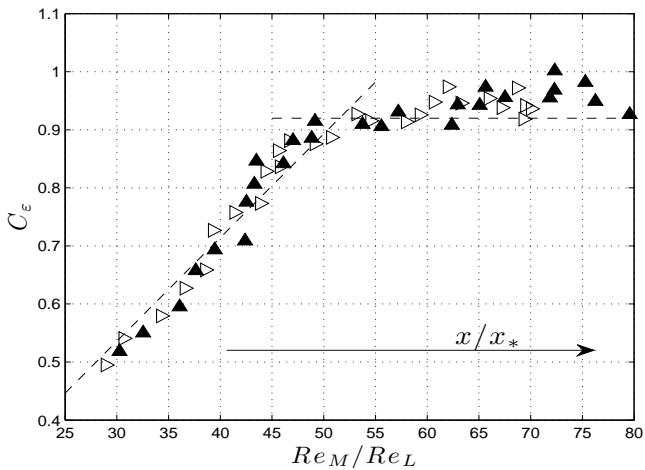}
\caption{\label{fig4}  Normalised energy dissipation $C_{\varepsilon}$ versus the Reynolds number ratio $Re_M/Re_{L}$ of RG60-generated turbulence. }
\end{figure}

% paragraph 10
Equation (\ref{eq1}) may appear to clash with the fact that $C_{\varepsilon}$ is approximately independent of both $x$ and $Re_M$ in the case of RG60 at distances greater than about 1.5m from that grid in a wind tunnel test section of exact same width as the present one (see Fig.  7 in \cite{VV11}). 
This is a distance greater than about $4x_{*}$ from the grid because $x_{*} \approx 0.36\mathrm{m}$ for RG60. 
However, (\ref{eq1}) has so far been established for decaying turbulence originating from RG230 and FSG up to downstream distances of less than about $1.5x_{*}$ ($x_{*}$ takes much greater values for these grids, see table \ref{grids}). 
It is therefore reasonable to investigate whether (\ref{eq1}) and its equivalent relation
$L/\lambda \sim Re_{M}^{1/2}$ hold at distances below a few multiples
of $x_{*}$ from the RG60 grid. 
In Fig. \ref{fig3} we plot $L/\lambda$ as a function of the local Reynolds number $Re_{\lambda}$ for RG60 at different levels of $Re_M$. 
We find that $L/\lambda \approx const$ in the region between $0.72 x_{*}$ and $2x_{*}$ (where $Re_{\lambda}$ takes the largest values) and that $L/\lambda$ and $Re_{\lambda}$ decay in exact proportion to each other (i.e. $L/\lambda \sim Re_{\lambda}$ which is equivalent to $C_{\varepsilon} = const$) at further downstream
distances, i.e.  where $x > 2x_{*}$. 
The region between $0.72 x_{*}$ and $2x_{*}$ corresponds to the 10 highest $Re_{\lambda}$ data points in Fig.  \ref{fig3}  for each $Re_{M}$. The $x$-independent (therefore $Re_{\lambda}$-independent) value of $L/\lambda$ in this region is
an increasing function of $Re_M$ as implied by (\ref{eq1}). 
Such $L/\lambda$ behaviour was previously reported only for FSGs \cite{VV11,MV10} and is now shown to be more general. 
Replotting the RG60 data so as to directly compare with (\ref{eq1}), we obtain Fig. \ref{fig4}. Equation (\ref{eq1}) is a fairly good representation of the data up to $Re_{M}/Re_{L} = 50$, i.e. in the turbulent decay region closest to the grid up to $x \approx 2x_*$. 
At streamwise distances larger than $2x_*$ where $Re_{M}/Re_{L}$ is larger than 50, $C_{\varepsilon}$ becomes approximately independent of both $x$ and $Re_M$ as already observed in earlier studies (e.g. \cite{VV11}).

% paragraph 11
Our measurements of decaying turbulence originating from RG115 were designed for a direct comparison with RG230 at equal $\sigma$ and $Re_{M} = 1.53 \!\times\! 10^5$ but different mesh size $M$. 
The data obtained from these measurements are reported in Figs. \ref{fig1} and \ref{fig3} and show that $L/\lambda$ and $C_{\varepsilon}$ take effectively same values for the two grids and that these values are consistent with $C_{\varepsilon}=f(Re_{M})/Re_{L}$ and constant $L/\lambda = \sqrt{f(Re_{M})/15}$ in the ranges of $x$ probed. However, $Re_M$ is too low for (\ref{eq1}) to hold.

% paragraph 12
The present data and those of \cite{VV11,MV10} conspire to form the conclusion
that, irrespective of the turbulence generating grid (Fig. \ref{figGrids}) and for high
enough $Re_M$,
\begin{equation}
\varepsilon \approx C_{1} \frac{U_{\infty} u^{2}}{L}\frac{M}{L}
\label{eq2}
\end{equation}
and equivalently $L/\lambda \approx \sqrt{C_{1} Re_{M}/15}$ are acceptable approximations in the non-equilibrium decay region $x_{peak} < x < x_{e}$ where $x_{e} \approx 2x_{*}$ for RG60 and $C_1$ is a dimensionless constant which only
depends on inlet/boundary geometry (type of fractal/regular grid,
$\sigma$, etc). 
We might expect $x_e$ to scale with $x_*$ for other grids as well, and the equilibrium dissipation scaling $\varepsilon = C_{2} u'^{3}/L$ (where $C_2$ is an inlet/boundary geometry-dependent dimensionless constant, see \cite{M&V2008,VV11PLA}) to be recovered at $x>x_{e}$ for other grids too. 
However, our RG115, RG230 and FSG data and those of  \cite{VV11,MV10} do not allow us to test these expectations, nor do they allow us to explore how $x_{e}/x_{*}$
may depend on inlet/boundary conditions.
RG230 and FSG, in particular, act as magnifying lenses which make the non-equilibrium region to be longer than the entire tunnel section's length. Equations (\ref{eq1}) and (\ref{eq2}), and more generally $C_{\varepsilon} = f(Re_{M})/Re_{L}$ which also covers lower values of $Re_M$, are approximately true in the non-equilibrium region irrespective of flow/turbulence profile details which differ from grid to grid. The FSGs are magnifying lenses with added capabilities for tailoring flow
and turbulence profiles which go beyond variations in $\sigma$. 

% paragraph 13
Finally, it is important to stress that the energy spectrum has a well-defined power-law shape over nearly two decades with exponent close to -5/3 at the closest point to the grid that we sampled in the non-equilibrium region (Fig. \ref{fig0}). This power-law region becomes progressively narrower with an exponent progressively further away from -5/3 as $x$ increases.  In the equilibrium region of RG60 where
$\varepsilon \sim u'^{3}/L$, the energy spectrum is far from Kolmogorov-shaped. This may just be a consequence of the low Reynolds numbers in the equilibrium region of our RG60 runs. But it is remarkable that a near-Kolmogorov power-law shaped energy spectrum does in fact appear well before the turbulence has had the time to reach equilibrium. A similar observation was made in \cite{Braza} where near-Kolmogorov power-law energy spectra were reported in a cylinder wake within one cylinder diameter from the cylinder.\\

We are grateful to Zellman Warhaft for stimulating discussions. P.C.V acknowledges the financial support from Funda\c{c}\~{a}o para a Ci\^{e}ncia e a Tecnologia (SFRH/BD/61223/2009, co-financed by POPH/FSE).


\bibliography{MyAIPbib}

\end{document}